\documentclass[12pt]{iopart}
\pdfoutput=1
\usepackage{graphicx}
\usepackage{mathtools}
\usepackage{todonotes}

\usepackage{xcolor}
\usepackage{subfigure}
\usepackage{textcomp}
\usepackage{cite}
\usepackage{float}
\usepackage{soul}
\usepackage{stackrel}
\usepackage{hyperref}
\definecolor{dgreen}{rgb}{0,0.7,0}

 \expandafter\let\csname equation*\endcsname\relax
 \expandafter\let\csname endequation*\endcsname\relax

\usepackage{amsmath,amssymb}

\pdfoutput=1
\usepackage{esint}
\usepackage{amsfonts}
\usepackage{color}
\definecolor{dgreen}{rgb}{0,0.7,0}

\begin{document}

\title[]{First-passage functionals for Ornstein Uhlenbeck process with stochastic resetting}

\author{Ashutosh Dubey and Arnab Pal}
\address{The Institute of Mathematical Sciences, CIT Campus, Taramani, Chennai 600113, India \& Homi Bhabha National Institute, Training School Complex, Anushakti Nagar, Mumbai 400094, India}
\ead{ashutoshrd@imsc.res.in;~arnabpal@imsc.res.in}
\vspace{10pt}
%\begin{indented}
%\item[]August 2017
%\end{indented}
%\date{today}

\begin{abstract}
We study the statistical properties of first-passage Brownian functionals (FPBFs) of an Ornstein-Uhlenbeck (OU) process in the presence of stochastic resetting. We consider a one dimensional set-up where the diffusing particle sets off from $x_0$ and resets to $x_R$ at a certain rate $r$. The particle diffuses in a harmonic potential (with strength $k$) which is centered around the origin. The center also serves as an absorbing boundary for the particle and we denote the first passage time of the particle to the center as $t_f$. In this set-up, we investigate the following functionals: (i) local time $T_{loc} = \int _0^{t_f}d \tau ~ \delta (x-x_R)$ i.e., the time a particle spends around $x_R$ until the first passage, (ii) occupation or residence time $T_{res} = \int _0^{t_f} d \tau ~\theta (x-x_R)$ i.e., the time a particle typically spends above $x_R$ until the first passage and (iii) the first passage time $t_f$ to the origin. We employ the Feynman-Kac formalism for renewal process to derive the analytical expression for the first moment of all the three FPBFs mentioned above. In particular, we find that resetting can either prolong or shorten the mean residence and first passage time depending on the system parameters. The transition between these two behaviors or phases can be characterized precisely in terms of optimal resetting rates, which interestingly undergo a continuous transition as we vary the trap stiffness $k$. We characterize this transition and identify the critical -parameter \& -coefficient for both the cases.  We also showcase other interesting interplay between the resetting rate and potential strength on the statistics of these observables. Our analytical results are in excellent agreement with the numerical simulations.
\end{abstract}

\section{Introduction}
\label{introduction}
Brownian functionals, in their various different forms, appear ubiquitously in many different fields such as statistical physics \cite{Majumdar2005,chandrasekhar1943,redner2001,comtet2005functionals}, stochastic processes \cite{yor1992some,Kac1949,nobile1985}, chemistry \cite{Agmon1,Agmon2}, finance \cite{dufresne1990}, and computer science \cite{Majumdar2005}. The Brownian functional over fixed time interval $[0,t]$ is defined as $V=\int_{0}^{t}U[x(\tau)]d\tau$, where $x(\tau)$ is a Brownian path and $U[x]$ is some specified function of the path, which depends on the quantity one wants to calculate \cite{Majumdar2005}. Often, we are also interested in an another class of Brownian functional namely the first passage Brownian functional (FPBF), which is defined over random interval of time $[0,t_f]$ so that 
\begin{align}
V = \int _{0}^{t_f} U[x(\tau)] d \tau~,
\label{fun-eq-1}
\end{align}
where $t_f$ is the first-passage time which is random and will be defined more precisely later. In here, we consider an Ornstein-Uhlenbeck (OU) process i.e., a Brownian particle in a harmonic potential $\frac{1}{2}kx^2$, where $k$ is the strength or stiffness of the potential. The potential is centered at the origin. The equation of motion for the position $x(\tau)$ of the particle can be written in terms of the overdamped Langevin equation
\begin{align}
\frac{dx(\tau)}{d\tau} =-kx + \xi(\tau),
\label{BM-lanevin}
\end{align}
where $\xi(\tau)$ is the Gaussian white noise of zero mean and correlations $\langle \xi(\tau)\xi(\tau^{\prime})\rangle=2D\delta(\tau-\tau^{\prime})$. Here, $D$ is the diffusion constant and for simplicity, we have set $D=1$ unless specified otherwise. In addition, we will assume that there is an absorbing boundary at the origin  and the process ends as soon as the particle hits the origin, and so, $t_f$ encodes the statistics of this time. Due to the absorption, naturally there is no equilibration for the particle in this confining potential. 

In literature, Brownian functionals appear in a wide variety of problems across different disciplines. For instance, an important studied Brownian functional is the local time which quantifies the amount of time spent by the Brownian particle in the neighborhood of a desired location \cite{Majumdar2005,Sabhapandit:2006,McKean1975,Knight,graph,SinghKundu2021,reactive,Majumdar-Comtet2002}.  Another important example of such Brownian functional is occupation or residence time, which specifies the total time spent by a diffusing particle in a certain domain. The celebrated L\'evy arc-sine law is an estimation of residence time for simple 1D diffusion process when considering the total time spent above the origin \cite{Majumdar2005,Levy,Louchard}. Over
the years, residence time statistics has been computed for simple diffusion \cite{Agmon1,Agmon2} diffusion in random and porous landscape \cite{Majumdar-Comtet2002,porousD}, heterogeneous medium \cite{hetero} or in a potential \cite{Sabhapandit:2006}, in confinements\cite{Grebenkov2007}, in active particle systems \cite{Singharc}, in molecular biology \cite{Agmoan} and in experiments \cite{expt-residence}. The list is not exhaustive but most of these Brownian functionals have been calculated for a fixed interval of time. 

Another important class of functional is the first passage Brownian functionals (FPBFs) which are measured upto a random interval of time $t_f$, which keeps a record of time when a process satisfies certain criterion e.g. a particle reaching a desired location or marking turnover of an enzymatic reaction. The field of first passage time itself is remarkably broad and there are a myriad of applications that span over many different topics. We refer to \cite{redner2001,persistence-m,fpt-book,metzler-book} for an extensive read of the subject. The other important FPBFs studied in the literature are area and extreme maximum which have applications in statistical physics and queuing theory \cite{kearney2005,kearney2007,kearney2014}. The FPBF has also been calculated in the context of snowmelt dynamics \cite{dubey1}, biopolymer translocation dynamics \cite{dubey2}, DNA breathing dynamics \cite{dubey3} and barrierless reactions \cite{dubey4}. There has been a renewed interest for studying functionals in the presence of stochastic resetting. 
Brownian functionals such as the simple first passage time \cite{Restart1,Restart1-2,Restart1-3,Evansrev2020,PalJphysA,optimal-bridge}; local time \cite{local-r} \& occupation time \cite{occup-r-1,occup-r-2} till fixed interval, and the same along with the area functional till first passage event \cite{Prashant22} have been calculated in the presence of stochastic resetting.

In this work, we focus on various first passage functionals for OU process in the presence of stochastic resetting. To the best of our knowledge, treatment for OU process with resetting has been limited to non-equilibrium properties such as steady states and relaxation \cite{Restart4,OU-neq-1,OU-neq-2}, large deviations of time-additive functions \cite{OU-neq-3}, estimating work fluctuations \cite{OU-neq-4},  and the first passage time \cite{RT-2}, but a comprehensive study for the local and occupation/residence time functionals appears to be missing. This work tries to fill some of this void and while doing so, it also reveals an optimization and resetting transition for the residence time similar to the first passage time. For completeness, we mention a few works along this direction but in the absence of resetting. The first passage time \cite{nobile1985}, area \cite{RicciardiSato88,Kearney21,FPA-OU-2} and local time \cite{Kundu2021LT} have been calculated in the absence of stochastic resetting. Such computation for OU functionals has been normally invoked in biological systems particularly in the modeling of neuronal activity as mentioned in the reference \cite{nobile1985}. 

The central goal here is to study the statistical properties of FPBF-s for OU process in presence of resetting. We consider that the particle is reset to a location $x_R>0$ at a rate $r$. Under this assumption, motion of the particle in microscopic time step $\Delta t$ can be updated in the following way
% Stochastic resetting is a renewal process where the dynamics is repeated after some fixed or random amount of time. The modified dynamics will be the stochastic resetting to some position $x_R$ in addition to the dynamics governed by the Eq. \eqref{BM-lanevin}. More precisely, at time $t$, to update the location of particle $x(t)$ in time $dt$ has following two options: particle resets to $x_R$ with probability $rdt$ and with complimentary probability $(1-rdt)$, where the evolution of particle is governed by \eqref{BM-lanevin}. Hence, we have 
\begin{align}
x(t+\Delta t)=
\begin{cases}
&  x(t) -kx(t)\Delta t + \xi (t) \Delta t, ~~~~~~\text{with prob } (1-r \Delta t), \\
&  x_R,  ~~~~~~~~~~~~~~~~~~~~~~~~~~~~~~~~~~\text{with prob } r \Delta t.
\end{cases}
\label{update}
\end{align}
We consider the following observables in this study:
\begin{enumerate}
\item \textit{Local time}: Local time $T_{loc}$ refers to the amount of time spent by the particle in the neighbourhood of some desired position (say $x_{\ell}$) till the first passage time. For this quantity, $U(x) = \delta(x-x_{\ell})$ and hence $T_{loc}(x_\ell)=\int_0^{t_f}~d\tau~\delta[x(\tau)-x_\ell]$.  The $T_{loc}$ has explicitly been calculated near the resetting position in which case $x_{\ell} = x_R$.

\item \textit{Residence time}: The second functional is residence/occupation time, which estimates the cumulative time spent by the Brownian particle in a certain domain till the first passage time. In this case, $U(x) = \theta (x-x_R)$, where $\theta (x)$ is the Heaviside step function and the residence time takes the form $T_{res}(x_R)=\int_0^{t_f}~d\tau~\theta(x(\tau)-x_R)$.

\item \textit{First-passage time}: Finally, we consider the Brownian functional of the first-passage time itself. In this case $U(x)=1$ and hence the relevant functional $V=\int_{0}^{t_f}1~d\tau=t_f$ is the first passage time.
\end{enumerate}

\noindent
In this manuscript, we have illustrated, in detail, the effect of resetting and potential strength on the first moment of the above-mentioned FPBFs for the OU process. We find that the mean local time monotonically increases as a function of the resetting rate $r$. Furthermore, the mean residence time and the first passage time display both monotonic and non-monotonic behavior as a function of the resetting rate $r$. For non-monotonic case, it is generically found that resetting renders these times minima. The optimal resetting rates are shown to undergo a continuous phase transition as a function of the potential stiffness. We identify the critical strengths and characterize them using the generic framework of first passage under resetting wherever applicable. The manuscript is organized as follows. In section \ref{backward}, we revisit Feynman-Kac formalism and derive a backward differential equation for the moment generating function of an arbitrary functional. Utilizing this backward differential equation, we investigate the statistics of the local time in Sec. \ref{local}, residence time in Sec. \ref{residence}, and the first passage time in Sec. \ref{FPT}. We conclude our manuscript with a brief summary and future outlook in Sec. \ref{conclusion}.

\section{General formulation}
\label{backward}

This section will sketch out the essential steps 
to compute the first passage OU functionals in the presence of stochastic resetting. To this end, we will follow the Feynman-Kac formalism \cite{Majumdar2005,Prashant22}. Since we are focusing on the first passage functionals, the first step will be to derive a backward master equation for the moment generating function which is  defined as
\begin{align}
Q(p,x_0) &=\int _{0}^{\infty} dV ~e^{-p V}~P_R(V,x_0) \\
&=\langle e^{-p \int _{0}^{t_f} U[x(\tau)] d \tau} \rangle,
\label{def-Q}
\end{align}
where the average $\langle .. \rangle$ in Eq. \refeq{def-Q} represents the averaging over all such paths which start at some point $x_0$ at time $\tau=0$ and terminate at the origin at time $\tau=t_f$ in the presence of repeated resetting to $x_R$. There are three components of stochasticity in this problem (i) intrinsic thermal noise in the OU process, (ii) temporal randomness due to resetting, and (iii) randomness incurred to the final observation time $\tau=t_f$ due to the absorbing boundary. The moments of first passage Brownian functional $V$ can be derived directly from the moment generating function i.e.,
\begin{align}
\langle V^m \rangle = (-1)^m \left(\frac{\partial^m Q(p,x_0)}{\partial p^m}\right) \bigg|_{p \to 0}.
\label{moms-Q}
\end{align} 
 
To proceed further, consider one realization of a stochastic path $\{ x(\tau);0 \leq \tau \leq t_f \}$ and split that into two parts: (i) the first interval $[0,\Delta t]$ and (ii) the rest interval $[\Delta t,t_f]$ with $\Delta t \to 0$. In the first part, the Brownian path propagates from $x_0$ at $\tau=0$ to $x_0'=x_0 + \Delta x$ in time $\Delta t$. In the rest interval, the process continues until the particle, which is at $x_0'$ at time $\Delta t$, is absorbed at the origin at a later time $t_f$, possibly undergoing many resetting events. This path-decomposition leads us to break the integration in Eq. \eqref{def-Q} as $\int _{0}^{t_f} = \int _{0}^{\Delta t}+\int _{\Delta t}^{t_f}$ and in the limit $\Delta t \to 0$, we get
\begin{align}
Q(p,x_0) &= \langle e^{-p U(x_0) \Delta t} ~e^{-p \int _{\Delta t}^{t_f} U[x(\tau)] d \tau} \rangle, \nonumber\\ 
& =\langle e^{-p U(x_0) \Delta t} Q(p, x_0') \rangle.
\label{def-Q1}
\end{align}
To evaluate $x_0'$, note that the Brownian particle 
can reset to $x_R$ with probability $r \Delta t$ in an infinitesimal time $\Delta t$ or it can keep evolving according to Eq. \eqref{update}. Putting these two contributions together, using the noise properties  $\langle \xi  (0) \rangle $ and $\langle \xi ^2 (0) \rangle $, and
% starts from $x_0$ and can either move to $x_0 + \Delta x$ with probability $(1-r \Delta t)$ or r. Thus,
% \begin{align}
% x_0' &=x_0 -kx_0(t)\Delta t + \xi (t) \Delta t, ~~~~\text{with probability }(1-r \Delta t) , \nonumber \\
% & = x_R, ~~~~~~~~~~~~~~~~~~~~~~~~~~~~~~\text{with probability }r \Delta t
% \label{update-resetting}
% \end{align}
% We now use Eq. \eqref{update-resetting} in Eq. \eqref{def-Q1} along with the noise properties $\langle \xi  (0) \rangle $ and $\langle \xi ^2 (0) \rangle $ as $\Delta t \to 0$. 
keeping only the leading order terms in $\Delta t$, we obtain the following backward equation in the presence of resetting
\begin{align}
\frac{\partial ^2 Q(p,x_0)}{\partial x_0^2} - kx_0\frac{\partial Q(p,x_0)}{\partial x_0}- p  U(x_0) Q(p,x_0)-r Q(p,x_0)+r Q(p,x_R) = 0. 
\label{bfp}
\end{align}
Eq. \eqref{bfp}, valid in the domain $x_0\in[0,\infty]$, is the key equation of this work. This is moreover supplemented with the following boundary conditions
\begin{align}
& Q(p, x_0 \to 0^+) =1, ~\label{An-eq-2} \\
& Q(p, x_0 \to \infty) < \infty.
\label{An-eq-3}
\end{align}
The first boundary condition is obvious since the particle, starting close to the origin i.e., $x_0\rightarrow 0$, will immediately be absorbed yielding FPT $t_f\rightarrow0$. Therefore, substituting $V =\int_0^{t_f}~d\tau~U[x(\tau)] \to 0$ in Eq. \eqref{def-Q} results in  the first boundary condition \eqref{An-eq-2}. The second boundary condition can be explained in the following way. Consider a situation when the particle had started from $x_0\rightarrow\infty$. Nonetheless, the next resetting move will bring it back to $x_R$ in a finite time span. Therefore, the FPT $t_f$ should be generally finite for any non-zero and finite value of the resetting rate $r$. Hence, the functional $V$ should remain finite for the fixed value of resetting position $x_R$ even when the initial position $x_0\rightarrow \infty$. This leads to the second boundary condition \eqref{An-eq-3}.
\newline \indent 
Moving forward, the steps are as follows.
Depending on the functional $V[x(\tau)]$ of interest, one first solves the corresponding backward differential Eq. \eqref{bfp} with the appropriate boundary conditions. This gives us the moment generating function $Q(p,x_0)$ which then leads to the moments of the functional using Eq.~\refeq{moms-Q}. We study each of the aforementioned functionals in the subsequent sections.

\section{Local time}
\label{local}
The local time density that a particle spents at the position $x_\ell$ until the first-passage time $t_f$ is given by
\begin{align}
T_{loc}(x_\ell)=\int_{0}^{t_f}d\tau ~\delta[x(\tau)-x_\ell]~.
\label{local-reset}
\end{align}
%\begin{alignat}{1}
%T_{loc}(x_\ell)=\int_{0}^{t_f}d\tau ~\delta[x(\tau)-x_\ell]~.
%\label{local-reset}
%\end{alignat}
In Eq. (12) the delta function is defined within the limiting sense as below:
\begin{align}
T_{loc}(x_\ell)= \lim_{\epsilon \to 0}\frac{T_{2\epsilon}(x_\ell)}{2\epsilon},~~~\text{where}~~T_{2\epsilon}(x_\ell)=\int_0^{t_f} d \tau ~[\theta(x(\tau)-x_\ell-\epsilon)-\theta(x(\tau)-x_{\ell}+\epsilon)]. \label{new-ps-eq-e1}
\end{align}
 Here, $T_{2\epsilon}(x_\ell)$   measure of the total time spent by the particle inside the box $[x_\ell-\epsilon,x_\ell+\epsilon]$ till the first passage event. Thus, the normalization condition can be understood as $\int_0 ^{\infty} T_{loc}(x_\ell)dx_\ell=t_f$. Since we are interested in estimating the local time density near the resetting location $x_R$, we substitute $U(x)=\delta (x-x_R)$ into the backward Eq. \eqref{bfp} to obtain
\begin{align}
 \frac{\partial ^2 Q(p,x_0)}{\partial x_0^2} - kx_0\frac{\partial Q(p,x_0)}{\partial x_0} - p ~ \delta(x_0-x_R) Q(p,x_0)-r Q(p,x_0)+r Q(p,x_R) = 0. 
\label{BM-loc-eq-1}
\end{align}
For $x_0 \neq x_R$, the $\delta(x_0-x_R)$ term is eliminated and Eq. \eqref{BM-loc-eq-1} can be written as
 \begin{align}
\frac{\partial ^2 Q(p,x_0)}{\partial x_0^2} - kx_0\frac{\partial Q(p,x_0)}{\partial x_0} -r Q(p,x_0)+r Q(p,x_R) = 0. 
\label{BM-loc-eq-2}
\end{align}
The homogeneous part of the Eq. \eqref{BM-loc-eq-2}
\begin{align}
	\frac{\partial ^2 Q(p,x_0)}{\partial x_0^2} - kx_0\frac{\partial Q(p,x_0)}{\partial x_0} -r Q(p,x_0) = 0
	\label{BM-loc-eq-2.1}
	%\label{Appen-02}
\end{align}
can be solved by changing the variable from $Q(p,x_0)$ to $\tilde{Q}(p,x_0)$ using the following transformation
\begin{align}
	\tilde{Q}(p,x_0)=\exp\bigg(-\frac{1}{2}\int^{x_0}kx'dx'\bigg) Q(p,x_0)=\exp\bigg(-\frac{kx_0^2}{4} \bigg)Q(p,x_0).
	\label{BM-loc-eq-2.2}
	%\label{Appen-03}
\end{align}
Using \eqref{BM-loc-eq-2.2} in \eqref{BM-loc-eq-2.1}, we obtain the following differential equation for $\tilde{Q}(p,x_0)$
\begin{align}
	\frac{\partial ^2 \tilde{Q}(p,x_0)}{\partial x_0^2} +\bigg( -r+\frac{k}{2}-\frac{k^2x^2}{4} \bigg)\tilde{Q}(p,x_0)=0,
	\label{BM-loc-eq-2.3}
\end{align}
which has a standard solution \cite{NIST,eqs}
\begin{align}
	\tilde{Q}(p,x_0)=\mathcal{A} D_{-\frac{r}{k}}(\sqrt{k}x_0)+\mathcal{B} D_{\frac{r}{k}-1}(i\sqrt{k}x_0),
\end{align}
where $D_{\lambda}(z)$ is a parabolic cylinder or Weber function and $i=\sqrt{-1}$ is the imaginary number. Transforming back to the original variable, we have the following solutions
\begin{align}
Q(p,x_0) = \begin{cases}
&e^{kx_0^2/4}[\mathcal{A} D_{-\frac{r}{k}}(\sqrt{k}x_0)+\mathcal{B} D_{\frac{r}{k}-1}(i\sqrt{k}x_0)] +Q(p,x_R),~~~~\text{for }x_0 <x_R, \\
&e^{kx_0^2/4}[\mathcal{C} D_{-\frac{r}{k}}(\sqrt{k}x_0)+\mathcal{D} D_{\frac{r}{k}-1}(i\sqrt{k}x_0)] +Q(p,x_R),~~~~\text{for }x_0 >x_R.
\end{cases}
\label{BM-loc-eq-3}
\end{align}

There are four constants namely $\mathcal{A},~\mathcal{B},~\mathcal{C}$ and $\mathcal{D}$ in the equation above. To solve them, four boundary conditions on $Q(p,x_0)$ need to be specified. Out of these, two boundary conditions originate from the behaviour of $Q(p,x_0)$ as $x_0 \to 0^+$ and $x_0 \to \infty$ as mentioned in Eqs. \eqref{An-eq-2} and \eqref{An-eq-3} respectively. The other two conditions are essentially matching conditions -- one of them being 
the continuity of $Q(p,x_0)$ across the point $x_0=x_R$ i.e.,
\begin{align}
	&~~Q(p,x_R  ^+) = Q(p, x_R  ^-). \label{BM-loc-eq-5}
\end{align}
The other one is a discontinuity condition in the derivative of $Q(p,x_0)$ across the point $x_0=x_R$ which can be obtained by integrating Eq. \eqref{BM-loc-eq-1} from $x_0 = x_{R}-\delta$ to $x_0 = x_{R}+\delta$ and taking $\delta \to 0^+$ so that
\begin{align}
	 \left(\frac{\partial Q}{\partial x_0}\right)_{ x_R^+}-\left(\frac{\partial Q}{\partial x_0}\right)_{ x_R^-} =  p Q(p, x_R ).\label{BM-loc-eq-6}
\end{align}
The constant $\mathcal{D}$ becomes zero owing
to the boundary condition \eqref{An-eq-3}. Setting $x_0=x_R$, and after some manipulations, it can be seen from Eq. \eqref{BM-loc-eq-3}  that $Q(p,x_R)$ is completely characterized  
 by $\mathcal{C}$ which we find to be

\begin{figure}[t]
\includegraphics[scale=0.65]{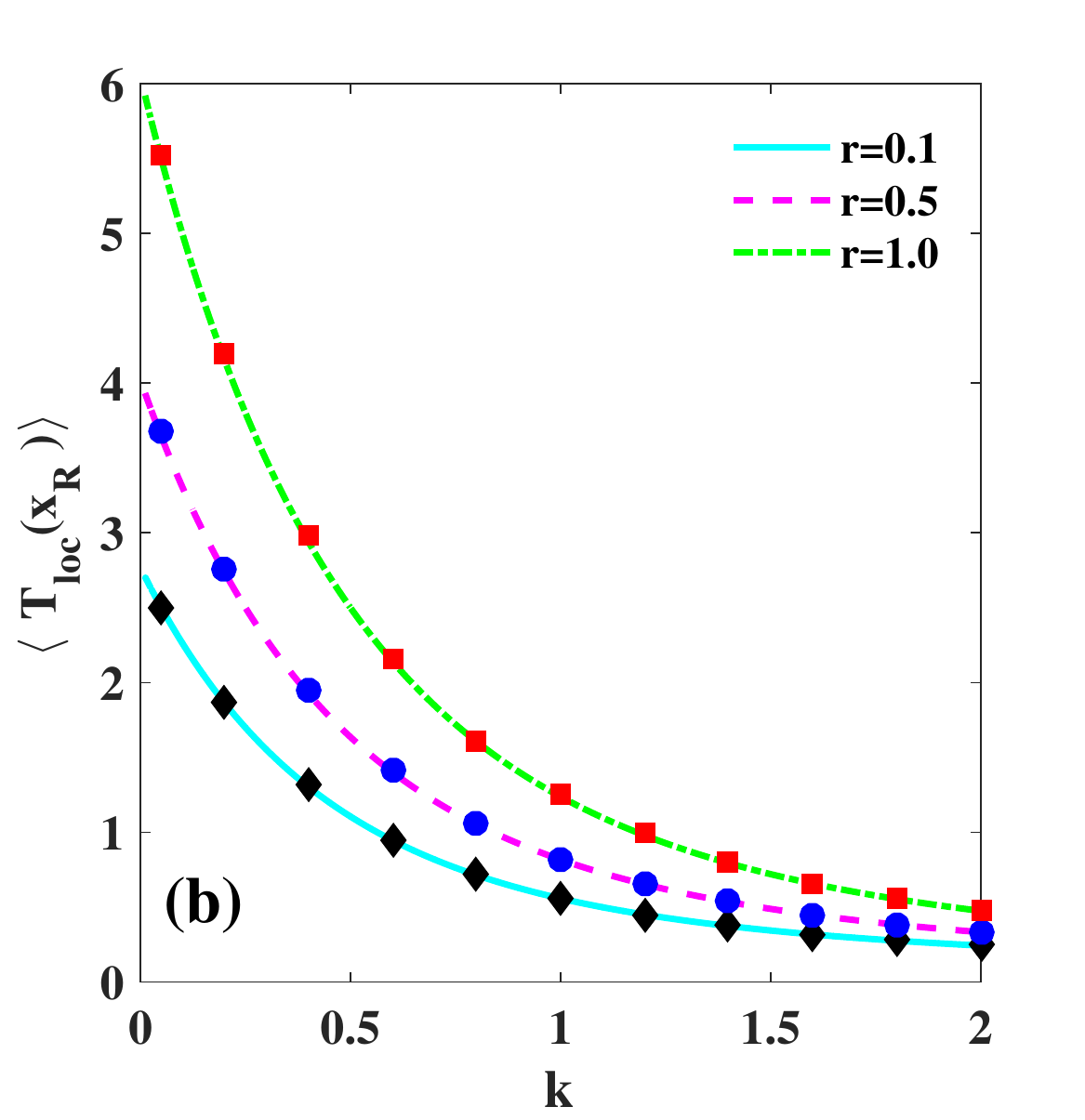}
\includegraphics[scale=0.65]{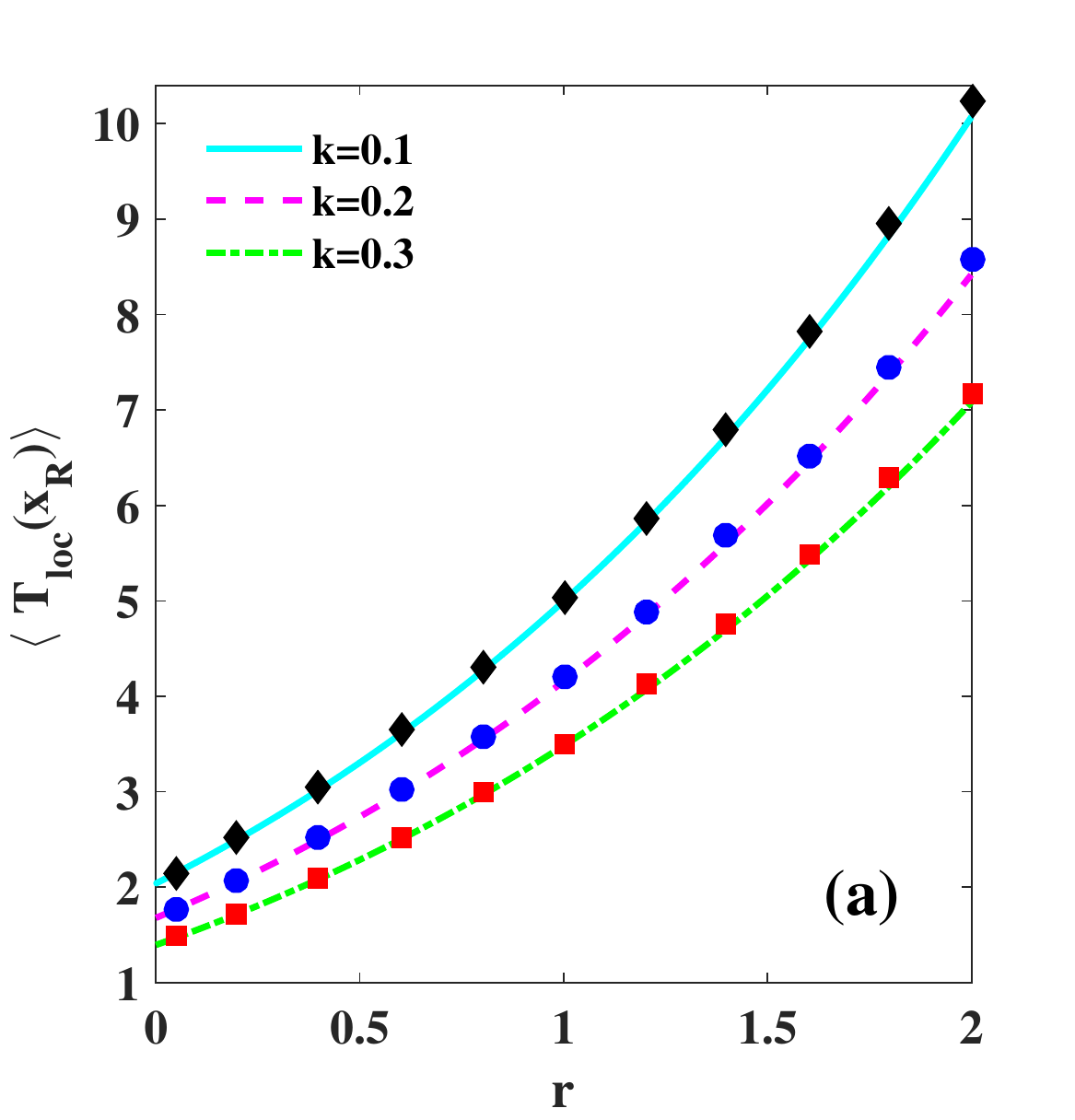}
\centering
\caption{Comparison of the first moment of $T_{loc}(x_R)$ for OU process with numerical simulation for different values of (a) stiffness constant  $k$ and (b) resetting rate $r$. Here, $x_0=x_R=2.5$. The lines and symbols are analytical and numerical results respectively.}  \label{local-moms-Fig}
\end{figure}
\begin{figure}[t]
	\includegraphics[scale=0.43]{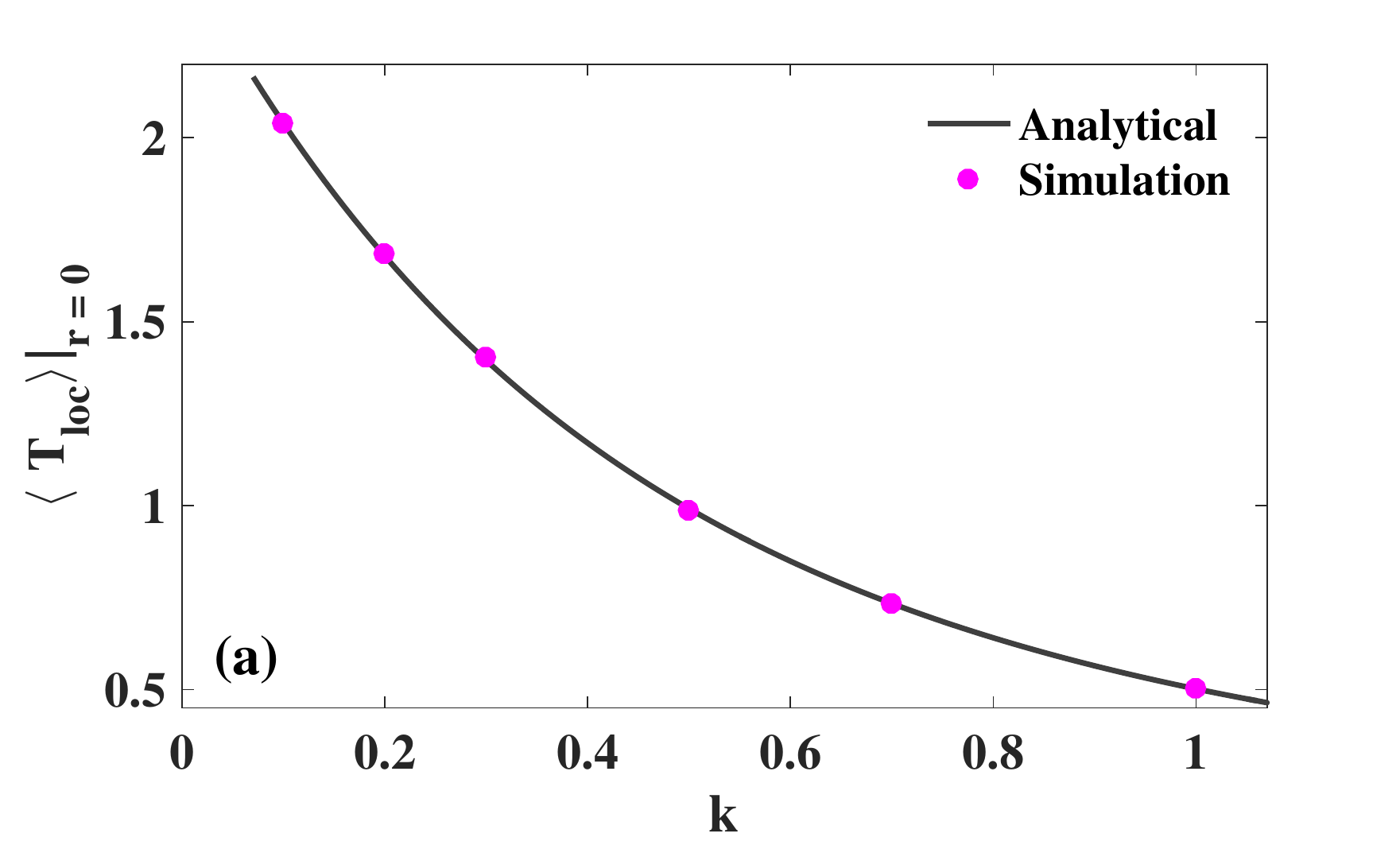}
	\includegraphics[scale=0.43]{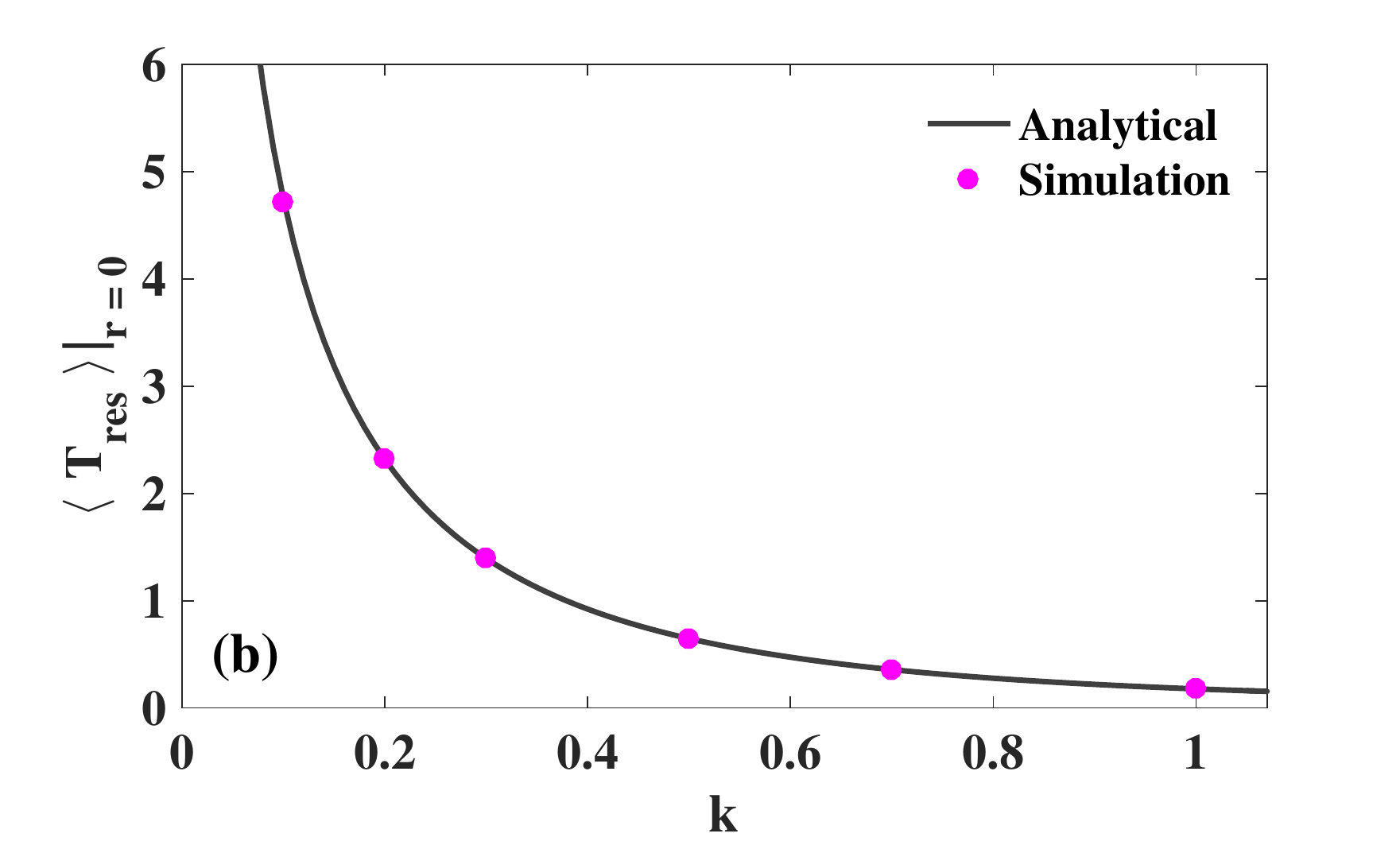}
	\centering
	\caption{Comparison of the theoretical expressions for the first moment of (a) local time density $\langle T_{loc}\rangle|_{r = 0}$ and (b) residence time $\langle T_{res}\rangle|_{r = 0}$ as a function of stiffness constant $k$ in the absence of resetting with numerical simulation. The lines and symbols are analytical and numerical simulations results respectively.  We have set: $x_0=2.5$. } \label{Resi-moms-Fig-r=0}
\end{figure}

 

\begin{align}
\mathcal{C}= \frac{\splitdfrac{pQ(p,x_R)e^{-kx_R^2/4}[D_{-\frac{r}{p}}(0)D_{\frac{r}{p}-1}(i\sqrt{k}x_R)-D_{\frac{r}{p}-1}(0)D_{-\frac{r}{p}}(\sqrt{k}x_R)]+[1-Q(p,x_R)]}{\times[i\sqrt{k}D_{-\frac{r}{p}}(\sqrt{k}x_R)D_{\frac{r}{p}}(i\sqrt{k}x_R)-\frac{r}{\sqrt{k}}D_{-(\frac{r}{p}+1)}(\sqrt{k}x_R)D_{\frac{r}{p}-1}(\sqrt{k}x_R)]}}{D_{-\frac{r}{p}}(0)[i\sqrt{k}D_{-\frac{r}{p}}(\sqrt{k}x_R)D_{\frac{r}{p}}(i\sqrt{k}x_R)-\frac{r}{\sqrt{k}}D_{-(\frac{r}{p}+1)}(\sqrt{k}x_R)D_{\frac{r}{p}-1}(i\sqrt{k}x_R)]},
\label{BM-loc-eq-7}
\end{align}
where we have used the following relations \cite{NIST,eqs}
\begin{align}
	&\frac{d^n}{dz^n}\bigg[ e^{z^2/4}D_{\nu}(z) \bigg]=(-1)^n(-\nu)_n 
	e^{z^2/4}D_{\nu-n}(z)\\
	&\frac{d^n}{dz^n}\bigg[ e^{-z^2/4}D_{\nu}(z) \bigg]=(-1)^n e^{-z^2/4}D_{\nu+n}(z)
\end{align}
Substituting Eq. \eqref{BM-loc-eq-7} into the second line of Eq. \eqref{BM-loc-eq-3} yields the complete solution for the local time moment generating function
\begin{align}
Q(p,x_R) = \left[1 + \frac{pe^{-kx_R^2/4}[D_{-\frac{r}{k}}(0)D_{\frac{r}{k}-1}(i\sqrt{k}x_R)-D_{\frac{r}{k}-1}(0)D_{-\frac{r}{k}}(\sqrt{k}x_R)]}{\frac{r}{\sqrt{k}}D_{-(\frac{r}{k}+1)}(\sqrt{k}x_R)D_{\frac{r}{k}-1}(i\sqrt{k}x_R)-i\sqrt{k}D_{-\frac{r}{k}}(\sqrt{k}x_R)D_{\frac{r}{k}}(i\sqrt{k}x_R)} \right]^{-1},
\label{BM-loc-eq-8}
\end{align}
The first moment of $T_{loc}$ can be calculated by inserting $Q(p,x_R)$ from the above into Eq. \eqref{moms-Q}. Skipping details, we find
\begin{align}
\langle T_{loc}(x_R) \rangle = \frac{2^{-(\frac{1}{2}+\frac{r}{2k})}e^{-kx_R^2/4}\sqrt{\pi k}[2^{\frac{r}{k}}\Gamma(\frac{1}{2}+\frac{r}{2k})D_{-\frac{r}{k}}(\sqrt{k}x_R)-\sqrt{2}\Gamma(1-\frac{r}{2k})D_{\frac{r}{k}-1}(i\sqrt{k}x_R)]}{\Gamma(1-\frac{r}{2k})\Gamma(\frac{1}{2}+\frac{r}{2k})[ikD_{-\frac{r}{k}}(\sqrt{k}x_R)D_{\frac{r}{k}}(i\sqrt{k}x_R)-rD_{-\frac{r}{k}-1}(\sqrt{k}x_R)D_{\frac{r}{p}-1}(i\sqrt{k}x_R)]},
\label{BM-loc-eq-9}
\end{align}
where $\Gamma(.)$ is the Euler gamma function. In the $r=0$ limit, the above Eq. \eqref{BM-loc-eq-9} takes the following form 
\begin{align}
    \langle T_{loc} \rangle\big|_{r=0} =\frac{2^{-1/2}e^{-kx_0^2/4}\sqrt{\pi k}[\Gamma(1/2)D_0(\sqrt{k}x_0) - \Gamma(1)D_{-1}(i\sqrt{k}x_0)]}{i\Gamma(1)\Gamma(1/2) k D_0(\sqrt{k}x_0)D_0(i\sqrt{k}x_0)},
    \label{lim_r=0_1}
\end{align}
which can be further simplified by noting
 $D_0(z) = e^{-z^2/4}$ and $D_{-1}(z) = \sqrt{\frac{\pi}{2}}e^{z^2/4}[1-\text{erf}(z/\sqrt{2})]$ to find
\begin{align}
\langle T_{loc} \rangle \big|_{r=0} =  \sqrt{\frac{\pi}{2k}}\exp\bigg[-\frac{kx_0^2}{2}\bigg] \text{erfi }\Big(\sqrt{\frac{k}{2}}x_0\Big),
\label{lim_r=0_2}
\end{align}
where $\text{erfi }(\cdot)$ is the imaginary error function defined by $\text{erfi }(z)=-i\text{erf}(iz)$ \cite{Kundu2021LT}. In Fig. \ref{local-moms-Fig} (a), we have plotted the first moment of local time density $\langle T_{loc}(x_R) \rangle$ as a function of resetting $r$ for different values of potential strength $k$. We found a monotonic increase in the mean local time as the  resetting rate is gradually increased. This is due to the fact that the particle will spend more time in the vicinity of $x_R$ if the frequency of resetting is higher. Fig. \ref{local-moms-Fig} (b) shows the monotonic decrease in $\langle T_{loc}(x_R) \rangle$ with $k$. As $k$ increases, the particle will have more drift towards the minimum of the potential, which is also the location of the absorbing boundary, and it gets absorbed more easily. As a result, the particle will spend lesser time around the resetting location $x_R$. At a sufficiently large value of $k$, $\langle T_{loc}(x_R) \rangle$ is almost in the same order for different $r$ since the trap is so strong that the particle is not able to diffuse much through the space and gets absorbed almost momentarily despite a possible number of resetting events. For completeness, we also verify the mean local time $\langle T_{loc} \rangle \big|_{r=0}$ in Fig. \ref{Resi-moms-Fig-r=0}a.  We compared our analytical results with the numerical simulations and found an excellent agreement.

\section{Residence time}
\label{residence}
In this section, we discuss the effect of resetting on the first moment of occupation or residence time $T_{res}=\int_0^{t_f}~d\tau~\theta(x(\tau)-x_R)$. As sketched in Sec. \ref{backward}, we need to solve the backward differential equation \eqref{bfp} with $U(x) = \theta(x-x_R)$ which gives
\begin{align}
\frac{\partial ^2 Q(p,x_0)}{\partial x_0^2} - kx_0\frac{\partial Q(p,x_0)}{\partial x_0}- p  \theta (x_0-x_R) Q(p,x_0)-r Q(p,x_0)+r Q(p,x_R) = 0. 
\label{BM-occ-eq-1}
\end{align}
Solving this equation separately for two regions $x_0 \geq x_R$ and  $x_0 < x_R$, we get
\begin{align}
Q(p,x_0) = \begin{cases}
&e^{kx_0^2/4}[\mathcal{C}_1D_{-\frac{r}{k}}(\sqrt{k}x_0) +\mathcal{C}_2 D_{\frac{r}{k}-1}(i\sqrt{k}x_0) ] +Q(p,x_R),~~~~~~~~~\text{for }x_0 <x_R,\\
&e^{kx_0^2/4}[\mathcal{C}_3 D_{-\frac{(r+p)}{k}}(\sqrt{k}x_0)+\mathcal{C}_4 D_{\frac{(r+p)}{k}-1}(i\sqrt{k}x_0)] +\frac{r}{r+p}Q(p,x_R),~~\text{for }x_0 >x_R .
\end{cases}
\label{BM-occ-eq-2}
\end{align}
As before, the four constants $\mathcal{C}_1,~\mathcal{C}_2,~\mathcal{C}_3$ and $\mathcal{C}_4$ (which are functions of $x_R$ and $p$, but are independent of $x_0$)
can be evaluated using the boundary conditions mentioned in Eqs. \eqref{An-eq-2} and \eqref{An-eq-3} along with the continuity conditions for $Q(p,x_0)$ and its derivative across $x_0 = x_R$ namely
\begin{align}
& Q(p, x_R^+) = Q(p, x_R^-),\label{BM-occ-eq-5}\\
& \left( \frac{\partial Q}{\partial x_0} \right)_{x_R^+}=\left( \frac{\partial Q}{\partial x_0} \right)_{x_R^-}.\label{BM-occ-eq-6}
\end{align}
With these conditions, all the four constants can be calculated directly from Eq. \eqref{BM-occ-eq-2}. As we are interested in $x_0=x_R$, we need only  $\mathcal{C}_3$ and  $\mathcal{C}_4$. However, the boundary condition \eqref{An-eq-3} demands $\mathcal{C}_4=0$ and we are left with
\begin{align}
\mathcal{C}_3=
\frac{\splitfrac{[1-Q(p,x_R)][r D_{-(\frac{r}{k}+1)}(\sqrt{k}x_R)D_{(\frac{r}{k}-1)}(i\sqrt{k}x_R)-ik D_{-\frac{r}{k}}(\sqrt{k}x_R)D_{\frac{r}{k}}(i\sqrt{k}x_R)]}{+\frac{p}{r+p}e^{-k x_R^2/4}Q(p,x_R)[ik D_{-\frac{r}{k}}(0)D_{\frac{r}{k}}(i\sqrt{k}x_R)-r D_{\frac{r}{k}-1}(0)D_{-(\frac{r}{k}+1)}(\sqrt{k}x_R)]}}{\splitfrac{(r+p)D_{-(\frac{r+p}{k}+1)}(\sqrt{k}x_R)[D_{-\frac{r}{k}}(0)D_{\frac{r}{k}-1}(i\sqrt{k}x_R)-D_{\frac{r}{k}-1}(0)D_{-\frac{r}{k}}(\sqrt{k}x_R)]}{+D_{-\frac{r+p}{k}}(\sqrt{k}x_R)[ik D_{\frac{r}{k}}(i\sqrt{k}x_R)D_{-\frac{r}{k}}(0)-r D_{\frac{r}{k}-1}(0)D_{-(\frac{r}{k}+1)}(\sqrt{k}x_R)]}}.\label{BM-occ-eq-7}
\end{align}
On substituting \eqref{BM-occ-eq-7} in the second line of Eq. \eqref{BM-occ-eq-2}, we finally arrive at the following expression for the moment generating function of the residence time
\begin{align}
Q(p, x_R) = \left[ 1+\frac{pe^{-kx_R^2/4}D_{-\frac{r+p}{k}-1}(\sqrt{k}x_R)[ D_{-\frac{r}{k}}(0)D_{\frac{r}{k}-1}(i\sqrt{k}x_R)-D_{\frac{r}{k}-1}(0)D_{-\frac{r}{k}}(\sqrt{k}x_R) ]}{D_{-\frac{r+p}{k}}(\sqrt{k}x_R)[ rD_{-\frac{r}{k}-1}(\sqrt{k}x_R)D_{\frac{r}{k}-1}(i\sqrt{k}x_R)-ikD_{-\frac{r}{k}}(\sqrt{k}x_R)D_{\frac{r}{k}}(i\sqrt{k}x_R) ]} \right]^{-1}.
\end{align}

\begin{figure}[t]
	\includegraphics[scale=0.43]{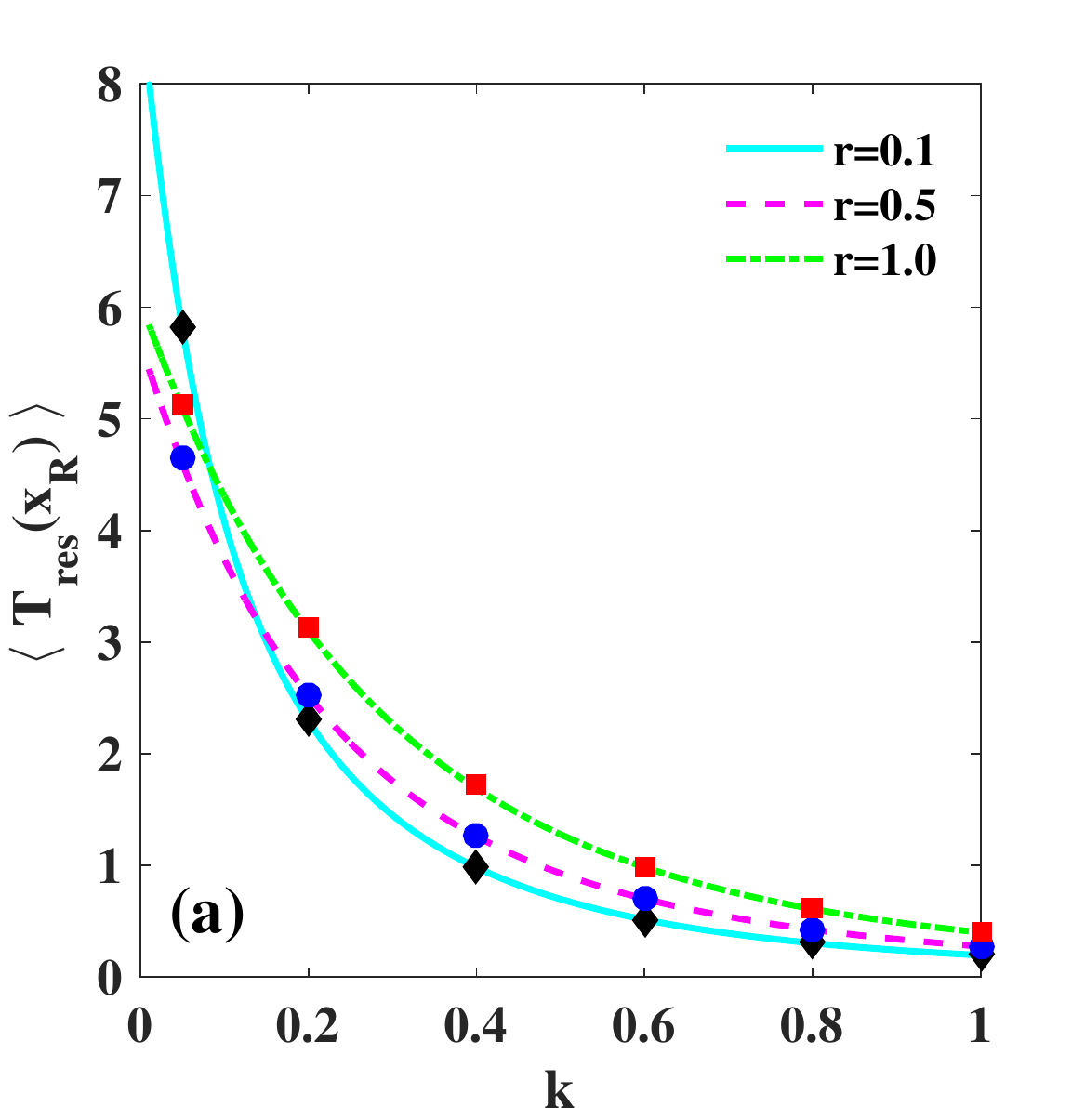}
	\includegraphics[scale=0.43]{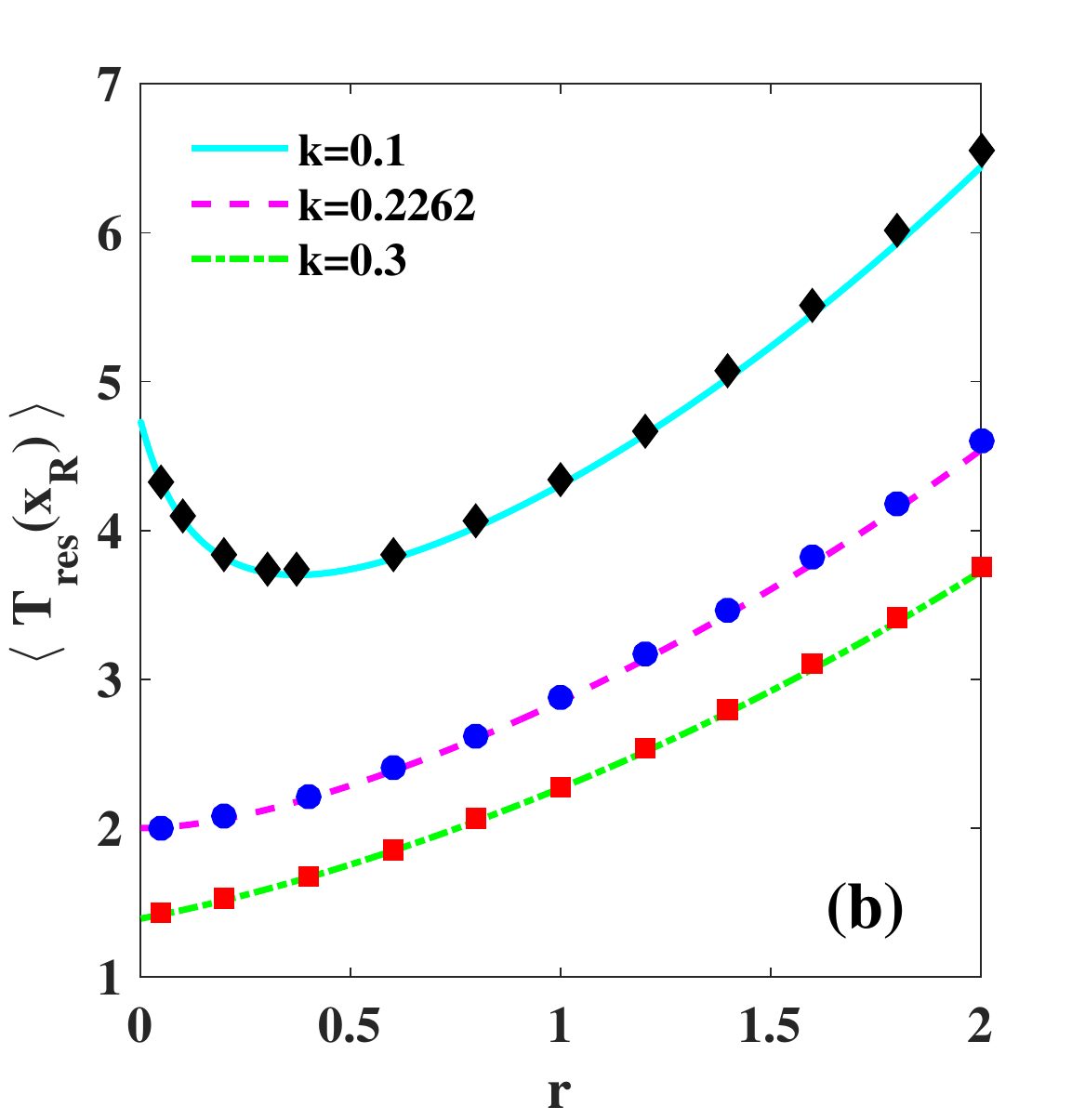}
 \includegraphics[scale=0.44]{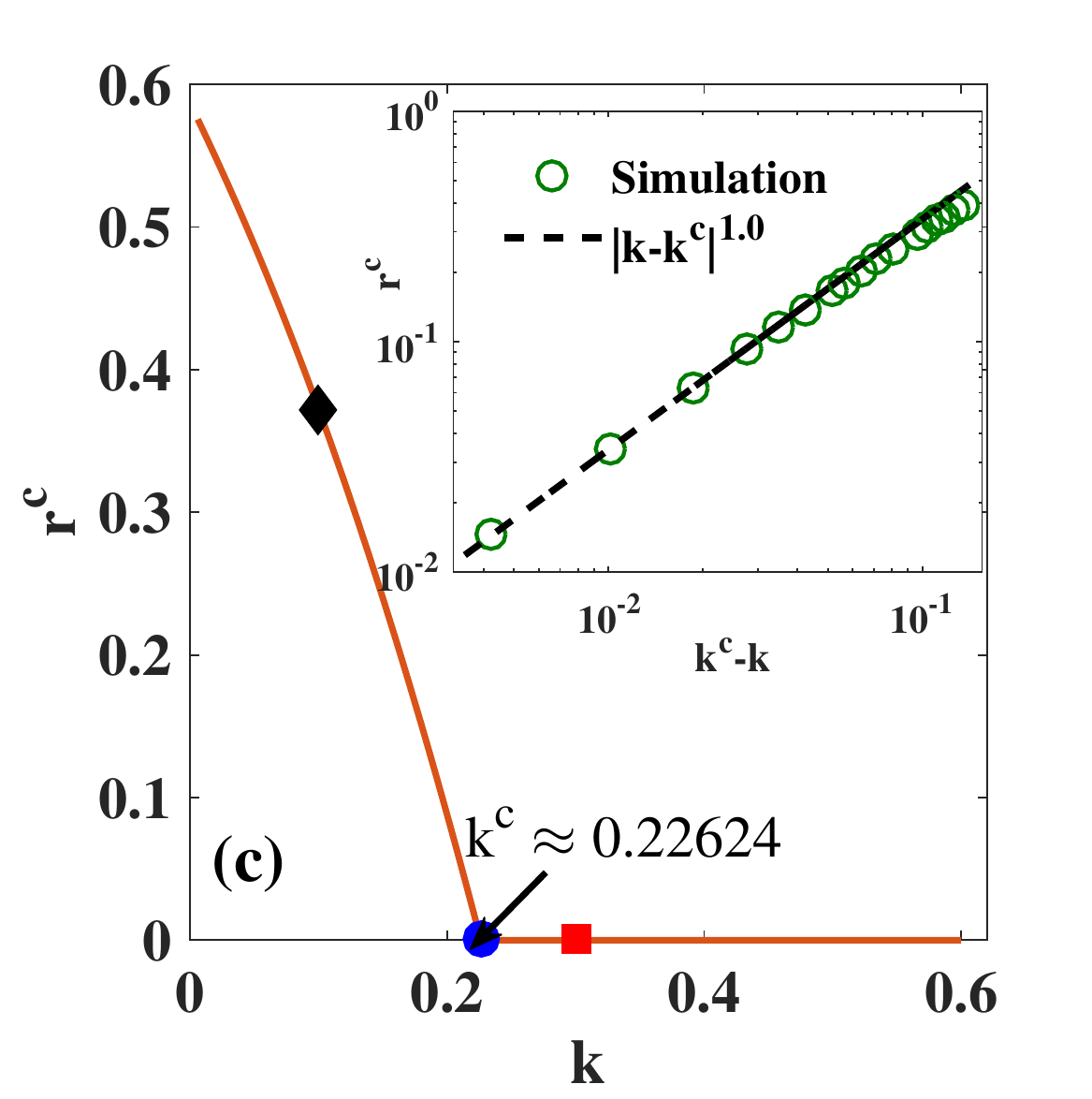}       
	\centering
	\caption{Comparison of the first moment of residence time $\langle T_{res}(x_R)\rangle$ for OU process with numerical simulation as a function of  (a) stiffness constant $k$ for different values of $r$ and, (b) resetting rate $r$ for different values of $k$. The lines and symbols are analytical and numerical simulations results respectively. (c) Continuous transition of optimal resetting rate $r^c$, which from a finite value transits to zero at the critical point $k^c = 0.22624$. The locus of $r^c$ (solid line) is obtained by numerically minimizing $\langle T_{res}(x_R)\rangle$ for different values of $k$. In particular, the diamond and square markers 
 depict $r^c>0$ and $r^c=0$ respectively obtained by minimizing $\langle T_{res}(x_R)\rangle$ from \ref{Resi-moms-Fig}(b) for two different values of $k$. Inset: $r^c$ near the critical point $k^c$ shows  a power law behavior. In the simulation, we have set: $x_0=x_R=2.5$ and $D=1$. } \label{Resi-moms-Fig}
\end{figure} 

\noindent
The mean residence time can be calculated using  Eq. \eqref{moms-Q} and reads
\begin{align}
\langle T_{res}(x_R) \rangle = \frac{i\sqrt{\pi}2^{-1-\frac{r}{2k}}e^{-kx_R^2/4}D_{-\frac{r}{k}-1}(\sqrt{k}x_R)[ 2\Gamma(1-\frac{r}{2k})D_{\frac{r}{k}-1}(i\sqrt{k}x_R)-2^{\frac{1}{2}+\frac{r}{k}}\Gamma(\frac{1}{2}+\frac{r}{2k})D_{-\frac{r}{k}}(\sqrt{k}x_R) ]
}{\Gamma(1-\frac{r}{2k})\Gamma(\frac{1}{2}+\frac{r}{2k})D_{-\frac{r}{k}}(\sqrt{k}x_R)[kD_{-\frac{r}{k}}(\sqrt{k}x_R)D_{\frac{r}{k}}(i\sqrt{k}x_R)+irD_{-\frac{r}{k}-1}(\sqrt{k}x_R)D_{\frac{r}{k}-1}(i\sqrt{k}x_R) ]}.\label{BM-occ-eq-9}
\end{align}
The above Eq. \eqref{BM-occ-eq-9} takes the following form in the absence of resetting
\begin{align}
\langle T_{res} \rangle \big|_{r=0} =  \frac{\pi}{2k}\bigg[1 - \text{erf}{\bigg(\sqrt{\frac{k}{2}}x_0\bigg)\bigg]} \text{erfi } \bigg(\sqrt{\frac{k}{2}}x_0\bigg).
\label{Limit_r=0}
\end{align}
Fig. \ref{Resi-moms-Fig} (a) shows the monotonic decrease of $\langle T_{res}(x_R) \rangle$ as a response to the strength of potential $k$. This is attributed to the fact that the Brownian particle can easily reach the target with an increase in $k$. For sufficiently larger values of $k$, $\langle T_{res}(x_R) \rangle$ saturates for different resetting rates. We have compared our analytical results (both for $r=0$ and $r>0$ case) with the numerical simulations and found an excellent agreement. 

Fig. \ref{Resi-moms-Fig} (b) showcases various interesting features of the mean residence time as a function of resetting rate. For large values of $k$, the mean residence time increases monotonically as a function of resetting rate since the potential is strong enough to bring the particle to the origin faster and thus with an increasing resetting rate the completion can only be delayed. As we keep decreasing the stiffness, an interesting regime appears where we see a 
a non-monotonic behaviour of $\langle T_{res}(x_R) \rangle$ with resetting rate $r$. We find that resetting first reduces $\langle T_{res}(x_R) \rangle$ and there is an optimal rate $r^c$ when $\langle T_{res}(x_R) \rangle$ is minimum. Following that, the mean residence time keeps increasing monotonically. A qualitative way of understanding this behavior is as follows: For the low values of $k$, the particle can easily diffuse through space and can easily spend more time above $x_R$. With resetting, the particle is brought back to $x_R$ and consequently it spends less time above $x_R$ and $\langle T_{res}(x_R) \rangle$ starts to decrease up to a critical resetting rate.  However, as $r$ is further increased, the particle is brought back to $x_R$ more frequently which increases the first passage time of the process, and in turn, the particle also gets to spend more time above $x_R$. 

It is evident from the above discussion that there is a critical strength $k^c$ above (and below) of which resetting has different impact on $\langle T_{res}(x_R) \rangle$. To capture this behavior, we have looked into the optimal rate $r^c$ as a function of $k$. For the large values of $k$, the mean residence time increases monotonically. Thus, the optimal resetting rate $r^c$ determined from $\frac{d\langle T_{res}(x_R) \rangle}{dr}|_{r=r^c}=0$ is always at zero in this case.
However, as we decrease $k$, a non-monotonic behavior emerges and we start observing a finite optimal rate $r^c>0$. Henceforth, there exists a critical potential strength $k^c$ where this transition occurs.  In Fig. \ref{Resi-moms-Fig} (c), we have shown the phase diagram in $(k,r^c)$ plane from which one can easily extract the value of critical strength $k^c$. For 
fixed values of $x_0=2.5$ and $D=1$, we find $k^c\approx0.22624$. The optimal rate  $r^c$ seems to exhibit a continuous phase transition alike behavior from $r^c>0$ to $r^c=0$. Any non-zero value of $k<k^c$ always renders in $r^c>0$ which essentially states that resetting will always minimize the mean residence time. However, for $k>k^c$, no benefits can be gained by the introduction of resetting. In other words, $\langle T_{res}(x_R) \rangle$ will increase monotonically as a function of resetting rate. At the proximity of the critical point $k^c$, one would expect a power law behavior $r^c \sim |k-k^c|^\beta$ similar to the thermodynamical second order phase transition. In the inset of Fig. \ref{Resi-moms-Fig} (c), we fit the numerical data with this ansatz to find the following critical coefficient $\beta=1$. This scaling behavior is similar to the one that we have found also for the MFPT $\langle t_f(r) \rangle$ as will be shown in the next section.

% . Here, the order parameter i.e. ORR $r^c$ exhibits the continuous phase transition, which reaches to zero at critical value $k^c\approx0.22624$ for fixed values of $x_0=2.5$ and $D=1$. The non-zero values of $r^c$ is the region in which the resetting helps and resetting will not help in the region in which $r^c$ is zero. The nonzero value of $r^c$ for the given value of $k$ minimizes the first moment of residence time. 

%\textcolor{red}{We need discussion for optimal rates $r^c$ at which mean residence time is minimum. This $r^c$ should be plotted as a function of $k$. This can be done numerically.} 

\section{First passage time}
\label{FPT}
In this section, we investigate the first moment of the first passage time $t_f$ of an OU particle, starting from $x_0$, to an absorbing target placed at the origin in the presence of stochastic resetting at $x_R>0$. To this end, we simply substitute $U(x)=1$ in Eq. \eqref{bfp} to find
\begin{align}
\frac{\partial ^2 Q(p,x_0)}{\partial x_0^2} - kx_0\frac{\partial Q(p,x_0)}{\partial x_0} -(p+r) Q(p,x_0) + r Q(p,x_R) = 0,
\label{OUP-FPT-eq-1}
\end{align}
with the boundary conditions \refeq{An-eq-2} and \refeq{An-eq-3}. The solution of above Eq. \refeq{OUP-FPT-eq-1} is
\begin{align}
Q(p,x_0)=e^{kx_0^2/4}[\mathcal{A}D_{-\frac{p+r}{k}}(\sqrt{k}x_0)+\mathcal{B}D_{\frac{p+r}{k}-1}(i\sqrt{k}x_0)]+\frac{r}{r+p}Q(p,x_R).
\label{OUP-FPT-eq-2}
\end{align}
Using boundary conditions \refeq{An-eq-2} and \refeq{An-eq-3} in Eq. \refeq{OUP-FPT-eq-2}, we obtain  $\mathcal{B}=0$ and
\begin{align}
	\mathcal{A}=\frac{1-\frac{r}{(r+p)}Q(p,x_R)}{D_{-\frac{p+r}{k}}(0)}.
	\label{OUP-FPT-eq-3}
\end{align}

\begin{figure}[t]
	\includegraphics[scale=0.65]{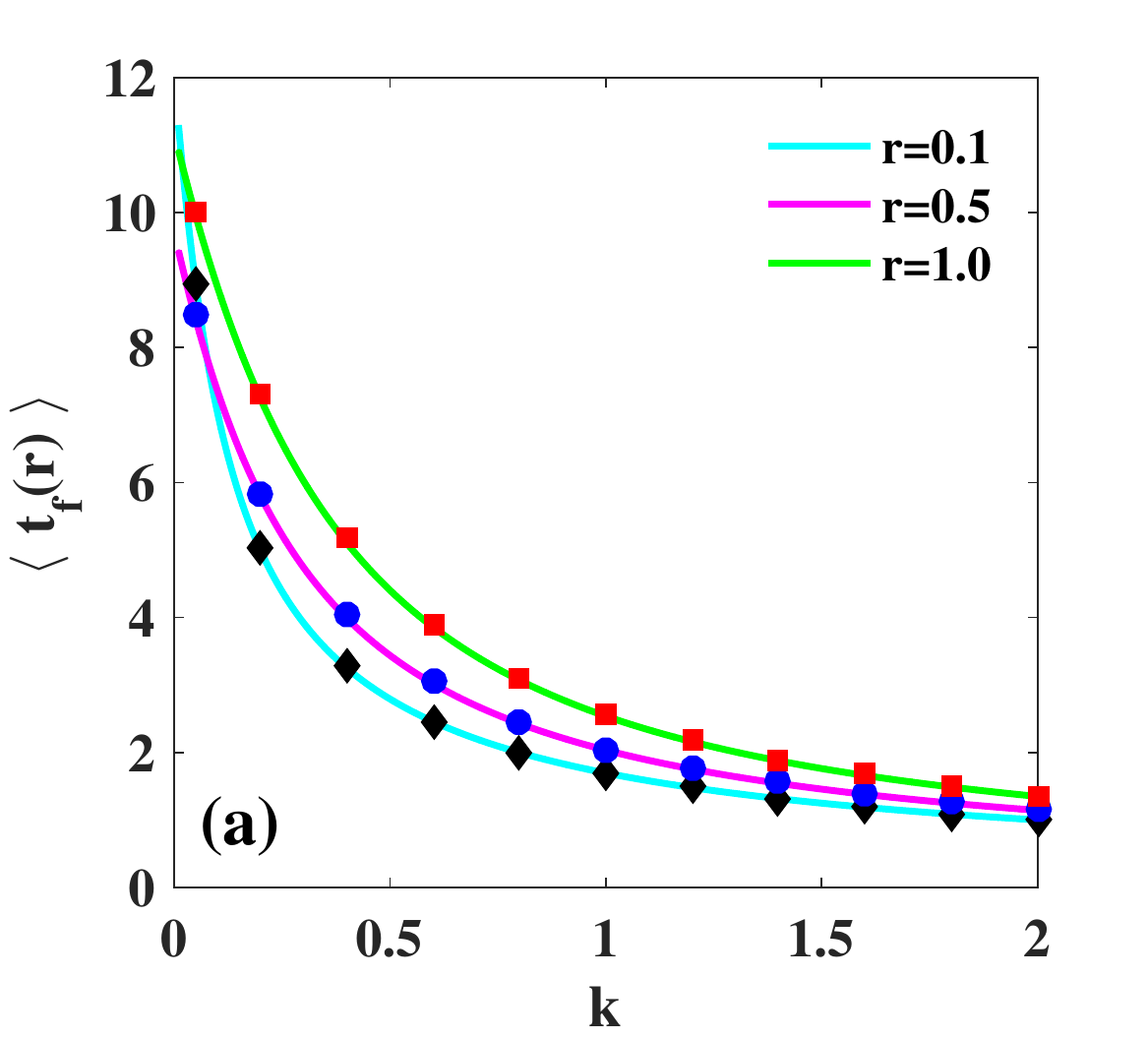}
	\includegraphics[scale=0.65]{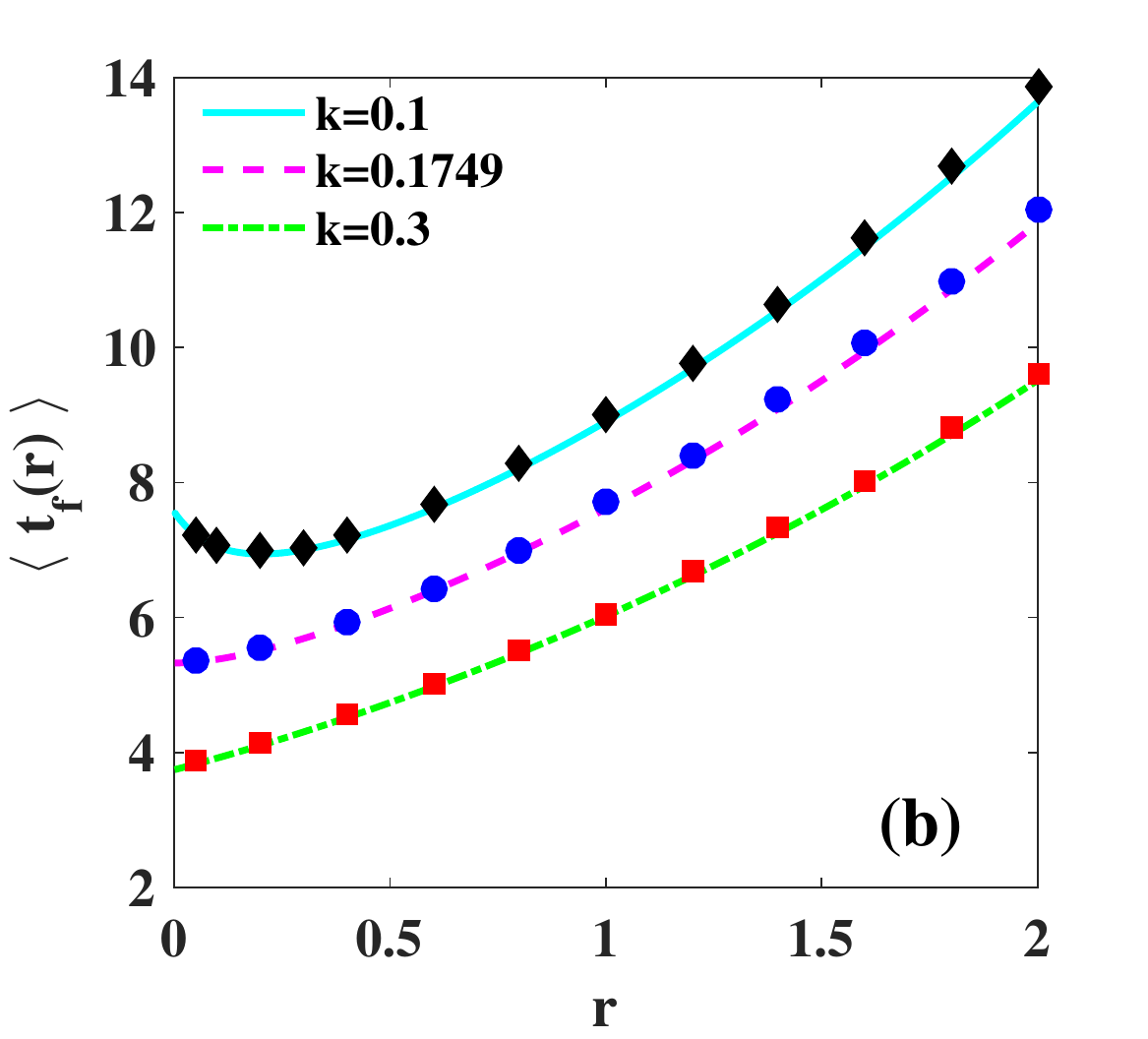}
    \includegraphics[scale=0.65]{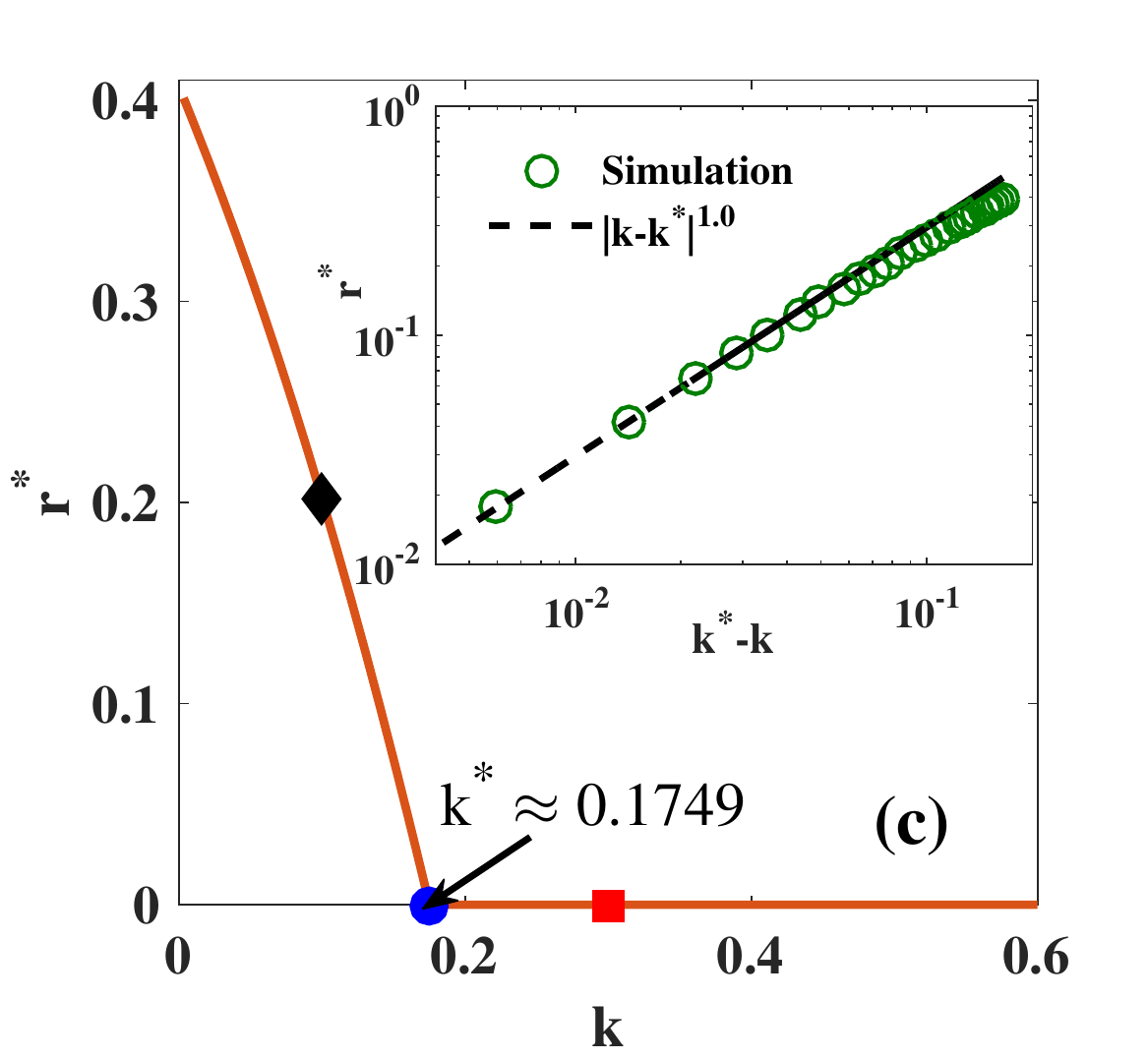}
    \includegraphics[scale=0.65]{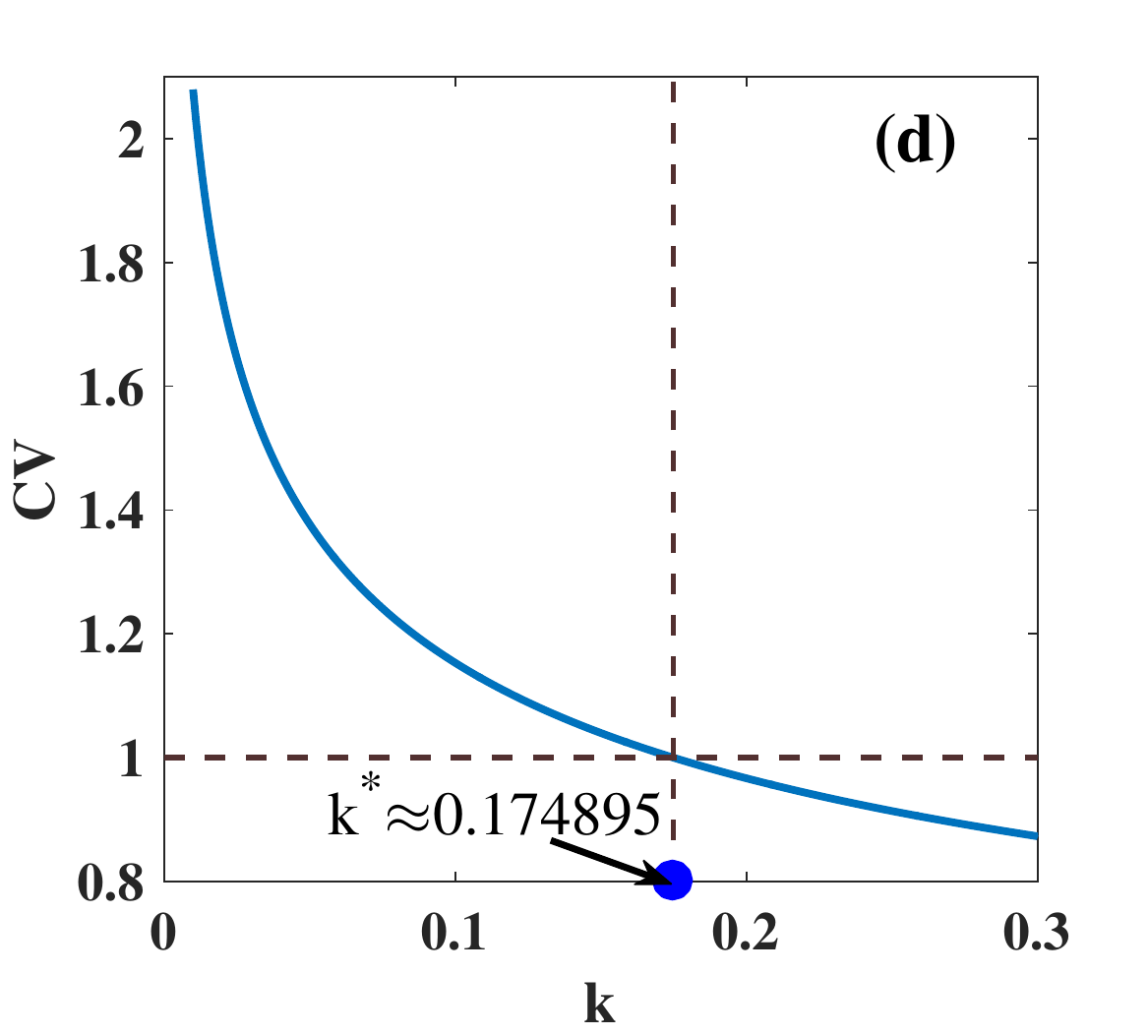}
	\centering
	\caption{Comparison of the MFPT $\langle t_f(r) \rangle$ for OU process with numerical simulations for different values of (a) resetting rate $r$ and (b) spring constant $k$. The lines and symbols are analytical and numerical simulation results respectively. (c) Second order like phase transition of the optimal resetting rate $r^*$ which continuously reaches to zero from a finite value at the critical point $k^* = 0.1749$. The locus of $r^*$ (solid line) is obtained by numerically minimizing $\langle t_f(r) \rangle$ for different values of $k$. In particular, the diamond and square markers 
 depict $r^*>0$ and $r^*=0$ respectively obtained by minimizing $\langle t_f(r)\rangle$ from \ref{FPT-moms-Fig}(b) for two different values of $k$. Inset: Optimal resetting rate $r^*$ close to the critical point $k^*$ shows a power law behavior $r^* \sim |k-k^*|^\alpha$ with $\alpha=1$. (d) Phase diagram for resetting transition and verification of the critical point $k^*$ from the $CV>1$ criterion. The $CV$ as a function of $k$ intersects unity (dashed horizontal line) exactly at the value $k^* = 0.174895$. For $k>k^*$ i.e., when $CV<1$, resetting is detrimental while for $k<k^*$ i.e., when $CV>1$, resetting always remains beneficial. This is in accordance with Fig. \ref{FPT-moms-Fig}(c). In the simulation, we have set: $x_0=x_R=2.5$ and $D=1$.}  \label{FPT-moms-Fig}
\end{figure} 

\noindent
Substituting the $\mathcal{A}$ and $\mathcal{B}$ in \refeq{OUP-FPT-eq-2} and setting $x_0=x_R$, we finally obtain
\begin{align}
Q(p,x_R)=\frac{(r+p)e^{kx_R^2/4} D_{-\frac{r+p}{k}}(\sqrt{k}x_R)}{D_{-\frac{r+p}{k}}(0)p+re^{kx_R^2/4}D_{-\frac{r+p}{k}}(\sqrt{k}x_R)}.
\label{OUP-FPT-eq-4}
\end{align}
Inserting $Q(p,x_R)$ from Eq. \eqref{OUP-FPT-eq-4} into Eq. \eqref{moms-Q}, we arrive at the following expression for the mean first passage time
\begin{align}
\langle t_f(r) \rangle = \frac{\sqrt{\pi}2^{-\frac{r}{2k}}e^{-kx_R^2/4}}{r\Gamma(\frac{r+k}{2k})D_{-\frac{r}{k}}(\sqrt{k}x_R)}-\frac{1}{r},
\label{OUP_MFPT}
\end{align}
which was also obtained in \cite{RT-2}. In the limit $k \rightarrow 0$, one can use the following identity \cite{NIST,eqs}
\begin{align}
D_{\nu}(z)=2^{\nu/2}\sqrt{\pi}e^{-z^2/4}\bigg[\frac{1}{\Gamma(\frac{1-\nu}{2})}F\bigg(-\frac{\nu}{2},\frac{1}{2};\frac{z^2}{2}\bigg) - \frac{\sqrt{2}z}{\Gamma(-\frac{\nu}{2})}F\bigg(\frac{1-\nu}{2},\frac{3}{2};\frac{z^2}{2}\bigg) \bigg],
\label{prop}
\end{align}
where $F(\alpha,\gamma;z) = \sum_{n=0}^{\infty}\frac{(\alpha)_n}{(\gamma)_n}\frac{z^n}{n!}$ is the Kummer function with $(\delta)_n = \beta(\beta + 1)....(\beta + n -1)$
 in \eqref{OUP_MFPT}, to find
\begin{align}
    \langle t_f(r) \rangle_{k\rightarrow 0}=\frac{1}{r}(e^{\sqrt{r}x_R}-1),
\end{align}
which was first obtained by Evans and Majumdar in the case of simple diffusion \cite{Restart1}. 

In Figs. \ref{FPT-moms-Fig} (a) and (b), we have plotted $\langle t_f(r)\rangle$ as a function of the strength of potential $k$ for different values of resetting rates $r$ and as a function of $r$ for different values of $k$ respectively. The solid/dashed lines represent the analytical results while the markers are the points from numerical simulations. An excellent agreement between them is found. The qualitative behaviour of $\langle t_f(r)\rangle$ is similar as $\langle T_{res}(x_R)\rangle$. At low values of $k$, $\langle t_f(r)\rangle$ increases with an increase in $r$ and becomes almost independent of the resetting rates when the potential is much steeper with higher $k$ [see Fig. \ref{FPT-moms-Fig} (a)]. On the other hand, Fig \ref{FPT-moms-Fig} (b) shows that at low values of $k$, where the drift is not sufficiently strong, $\langle t_f(r)\rangle$ decreases with an increase in $r$ and reaches some minimum value at the optimal resetting rate $r^*>0$ and then increases again. For sufficiently larger values of $k$, resetting always delays the completion rendering in $r^*=0$. Henceforth, there exists a critical value of potential strength namely $k^*$ below (above) which resetting is helpful (detrimental). To better understand this behavior, we have made a phase diagram, as done in the case of residence time, in the $(k,r^*)$ plane  -- see Fig. \ref{FPT-moms-Fig} (c) for fixed values of $D = 1$ and $x_0 = x_R = 2.5$. For this configuration, the optimal resetting rate $r^*$ undergoes a continuous transition from the $r^*>0$ to $r^*=0$ phase and there is a sharp critical point in the potential strength at $k^* = 0.1749$ where this behavioral transition occurs [see also \cite{RT-2} where this transition was analyzed numerically]. In a nutshell, we have
\begin{align}
    \begin{array}{l}
r^*\left\{ \begin{array}{lll}
=0 &  & \text{if ~~}k >k^*\text{ }\\
 & \text{ \ \ }\\
>0&  & \text{if~~ }k<k^*\text{ .}
\end{array}\right.\text{ }\end{array}
\label{second-order-transition}
\end{align}
The continuous transition of the optimal resetting rate compels us to believe that at the close proximity of the critical point $k^*$, the optimal resetting rate has a power law behavior $r^* \sim |k-k^*|^\alpha$, where $\alpha$ is the critical coefficient. Fitting with the numerical data, we find $\alpha=1$ -- see inset of Fig. \ref{FPT-moms-Fig}(c) and also in \cite{RT-2}. This is in accordance with the Landau theory of resetting transition where one can show that indeed $\alpha=1$ is a universal critical coefficient \cite{RT-0} in a very general scenario. This is quite striking since the power law behavior is reminiscent of the same 
observed, e.g., in liquid-gas or ferromagnetic systems where the critical coefficient
$\alpha$ is found to be universal.

Eq. \eqref{second-order-transition} can also be understood from a more general viewpoint laid out by the framework of first passage under resetting \cite{Restart7,Restart8} and a Landau like phase transition theory in stochastic resetting \cite{RT-0}. To see this, one can express
the MFPT $ \langle t_f(r) \rangle$ as a power series in terms of the restart rate $r$ near the transition and show how by utilizing the relations between
the coefficients, it is possible to predict the emergence of second-order like continuous transitions \cite{RT-0,RT-2}. Expanding $ \langle t_f(r) \rangle$ around $r=0$ results in
\begin{align}
   \langle t_f(r) \rangle = b_0 + b_1r + b_2 r^2 + \cdots
   \label{Taylor}
\end{align}
where $b_i-$s are the expansion coefficients. To identify the coefficients, we make use of the generic expression for MFPT under stochastic resetting for an arbitrary stochastic process given by  \cite{Restart7,Restart8}
\begin{align}
 \langle t_f(r) \rangle = \frac{1-\Tilde{t}_f(r)}{r\Tilde{t}_f(r)}   ,
\end{align}
where $\Tilde{t}_f(r) = \langle e^{-rt_f} \rangle$ is the moment generating function for the FPT $t_f$. Doing a Taylor
series expansion of $\langle t_f(r) \rangle$ and comparing term by term with Eq. \eqref{Taylor}, we can easily find the structure of the coefficients namely $b_0 = \langle t_f \rangle|_{r=0}$, $b_1 = [ -\frac{\langle t_f^2 \rangle}{2} +\langle t_f \rangle^2]\big|_{r=0}$, $b_2 =[ \frac{\langle t_f^3 \rangle}{6} + \langle t_f \rangle^3 - \langle t_f \rangle\langle t_f^2 \rangle]\big|_{r=0}$ and so on, notably $\langle t_f^n \rangle|_{r=0}$ is the $n$-th moment of FPT distribution for the resetting free process. For resetting to be beneficial in the small $r$ limit, one should have $b_1<0$ which after some rearrangement leads to the following universal criterion \cite{RT-0,Restart7,Restart8,inspection}
\begin{align}
    CV = \frac{\sigma(t_f)}{\langle t_f \rangle} \bigg|_{r=0} >1,
\end{align}
where $\sigma(t_f)$ stands for the standard deviation of $t_f$ and $CV$ is called the coefficient of variation. The $CV$ condition provides a phase diagram which allows us to scan the parameter space and identify the desired regions where in resetting can be beneficial. Within our set-up, we find
\begin{align}
  b_0=\frac{\exp\Big(-\frac{kx_R^2}{4}\Big)\bigg( 2D^{(1,0)}_0(\sqrt{k}x_R) - log(2)D_0(\sqrt{k}x_R) -D_0(\sqrt{k}x_R)\psi^{(0)}(1/2) \bigg)}{2k[D_0(\sqrt{k}x_R)]^2},
  \label{coef_b0}
\end{align}
and
\begin{align}
   b_1 = \frac{\splitfrac{\exp\Big(-\frac{kx_R^2}{4}\Big) \bigg[ [D_0(\sqrt{k}x_R)]^2\Big(2[log(2)]^2+4log(2)\psi^{(0)}(1/2)+2[\psi^{(0)}(1/2)]^2 -\pi^2 \Big) }{-8D_0(\sqrt{k}x_R)D^{(1,0)}_0(\sqrt{k}x_R)\Big(log(2)+\psi_0(1/2) \Big) +16[D^{(1,0)}_0(\sqrt{k}x_R)]^2 -8D_0(\sqrt{k}x_R)D^{(2,0)}_0(\sqrt{k}x_R) \bigg]}}{16k^2[D_0(\sqrt{k}x_R)]^3} ,
   \label{coef_b1}
\end{align}
where $D^{n,0}(\nu,z)$ is the $n^{th}$ partial derivative of parabolic cylinder function with respect to $\nu$ and $\psi^{0}(z)$ is the digamma function defined in terms of the gamma function $\Gamma(z)$ as $\psi^{(0)}(z) = \frac{\Gamma^\prime(z)}{\Gamma(z)}$ \cite{NIST,eqs}. Now substituting $b_0$ and $b_1$ from Eq. \eqref{coef_b0} and \eqref{coef_b1} into $ CV = \sqrt{1-\frac{2b_1}{b_0^2}}$ allows us to examine the span of the parameters e.g., the potential strength in our case by setting the condition $CV>1$. In Fig. \ref{FPT-moms-Fig} (d), we do this task by plotting CV as a function of $k$ for fixed $x_R$. From the intersection of $CV$ with unity, we can immediately identify the critical strength $k^* = 0.174895$. This is in perfect synchrony with the same obtained from the analysis of the optimization of $\langle t_f(r) \rangle$ as done around Eq. \eqref{second-order-transition}. It is also evident that for $k>k^*$, $CV$ becomes lesser than unity and thus resetting is only detrimental. On the other hand, $CV$ is greater than unity for  $k<k^*$, and thus resetting remains beneficial. Thus, in contrast to simple diffusion, where resetting always turns out to be a useful strategy \cite{Restart1}, a trade-off exists in the case of OU process \cite{RT-2}. The discussion made above exactly pinpoints how the potential strength $k$ plays a key role in this interplay.

\section{Conclusion}
\label{conclusion}
In recent times, study of stochastic processes that are further subject to resetting has gained immense interest in statistical physics \cite{Evansrev2020}. Resetting is a natural way to break detailed balance and thus resetting induced dynamics showcases a plethora of exciting non-equilibrium phenomena \cite{Restart1,Restart4,relaxation,eco-2,eco-3,eco-4,eco-5,eco-6,eco-7,eco-8}. In the context of search or completion of a process, resetting has been instrumental to curtail long waiting times and improve the efficiency \cite{Restart1,Evansrev2020,Restart7,inspection}. In this paper, we have analyzed various first passage functionals for an OU process subjected to stochastic resetting. Employing the celebrated Feynman-Kac formalism, we have computed the analytical forms for first moments of local time $T_{loc}$, residence time $T_{res}$ and first passage time $t_{f}$. While mean local time behaves monotonically as a function of the resetting rate, an intriguing feature is observed for the residence time. We have found that resetting can either shorten or lengthen the mean residence time and there exists an optimal resetting rate for which mean residence time can further be minimized. Furthermore, we can quantify the transition in terms of the  optimal resetting rate which can be interpreted as an order parameter in this problem. We show that the optimal resetting rate for the residence time undergoes a continuous transition as a function of the critical stiffness which we could exactly quantify. Moreover, we extracted the critical coefficient from the numerical analysis.

Finally, we study the mean first passage time of the OU process under resetting. Here too, resetting can either
accelerate or hinder the completion of the first passage process. The transition between these two states or phases is characterized by the behavioral change in the order parameter of the system namely the optimal restart rate which undergoes a second-order like continuous transition depending on the details of the system parameters. We quantify this transition via direct optimization of the mean first passage time and then provide a cross-verification of the same using the restart transition theory of optimal resetting rate \cite{RT-0}. Such resetting transition has been observed in various other systems such as diffusion in confinement \cite{Restart12} or channel \cite{channel}, diffusion in potential \cite{RT-1,RT-2,RT-4,RT-5,RT-6,RT-7,ray,Restart19}, random walks \cite{RW-1,records} and the universal applicability of the general condition \cite{Restart7,inspection} that dictates such transition is quite remarkable. 

Concluding, we have shown resetting induced optimization of the mean residence time and the mean first passage time in OU process that depends crucially on the potential strength. Various other intricate interplay between the potential strength and resetting has been showcased. It will be interesting to measure these observables using widely applicable optical trap experiments \cite{expt-1,expt-2}, calibrated with harmonic potential. Another interesting direction would be to understand the effect of various other non-exponential resetting protocols \cite{PalJphysA,Restart7,Restart11,power-law} on these observable statistics. In particular, resetting at a fixed interval of time or sharp resetting has turned out to be quite useful in global optimization of mean first passage time among all the strategies \cite{Restart7,Restart-Search3} with intriguing applications to operation research \cite{OR}, and nonlinear dynamics \cite{nonlinear-1}. These frontiers remain open as a prospective future direction.   

\section{Acknowledgment}
AD acknowledges support from the IMSc Post-Doctoral Scholarship. AP gratefully acknowledges research support from the Department of Science and Technology, India, SERB Start-up Research Grant Number SRG/2022/000080 and Department of Atomic Energy, India.

\appendix
\section{Details of numerical simulations}
\renewcommand{\theequation}{A.\arabic{equation}}
\setcounter{equation}{0}
The statistical properties of FPBFs studied here in this work can be numerically computed by integrating the overdamped Langevin equation in presence of resetting. The numerical scheme is as follows. Starting from an initial location $x_0$, position of the Brownian particle is updated 
using the following simple rules
\begin{align}
	x(t+\Delta t)=
	\begin{cases}
		&  x(t) -kx(t)\Delta t + \Gamma_0, ~~~~~~\text{with prob } (1-r \Delta t), \\
		&  x_R,  ~~~~~~~~~~~~~~~~~~~~~~~~~~~~~~~~~~\text{with prob } r \Delta t.
	\end{cases}
	\label{num_01}
\end{align}
Where, $\Gamma_0$ is a random number sampled from a Gaussian distribution with zero mean and width is given by $\langle \Gamma_0^2  \rangle=2Dt$ (here, $D=1$). The protocol \eqref{num_01} continues till the particle reaches very close to the origin within a prescribed tolerance level. We consider that to be a first passage event of the process. Computation of the three quantities discussed in the Sec. \ref{introduction} is done as follows:
\begin{enumerate}
	\item \textit{Local time}: Since we are interested in the local time in the vicinity of initial (or resetting) location $x_R$, we compute the time $T_{2\epsilon}(x_R)$ spent by the particle inside the box of length $[x_R-\epsilon,x_R+\epsilon]$. This process is observed till the particle reaches the origin and gets absorbed. We have chosen $\epsilon=0.1$. Finally, to compute the Local time density we devide the time $T_{2\epsilon}(x_R)$ by the box length as defined in the Eq. \eqref{new-ps-eq-e1}.
	\item \textit{Residence time}:
	We numerically evaluate the total time $T_{res}(x_R)$ spent by the Brownian particle in the region $x>x_R$ till the first passage event. 
	\item \textit{First passage time (FPT)}: To compute the FPT, we let the position evolve according to \eqref{num_01}. As soon as the particle is close to the origin within a tolerance level, we stop the process and record the total time spent so far as the FPT. 
\end{enumerate}
For averaging purpose, we repeat the same process for $10^5$ number of realizations.

\end{document}